# Propagation of solitons in thermal media with periodic nonlinearity

Yaroslav V. Kartashov,[1] Victor A. Vysloukh,[2] and Lluis Torner[1]

[1]ICFO-Institut de Ciencies Fotoniques, and Universitat Politecnica de Catalunya,
Mediterranean Technology Park, 08860 Castelldefels (Barcelona), Spain

[2]Universidad de las Americas – Puebla, 72820, Puebla, Mexico

We address the existence and properties of solitons in layered thermal media made of alternating focusing and defocusing layers. Such structures support robust bright solitons even if the averaged nonlinearity is defocusing. We show that non-oscillating solitons may form in any of the focusing domains, even in those located close to the sample edge, in contrast to uniform thermal media where light beams always oscillate when not launched exactly on the sample center. Stable multipole solitons may include more than four spots in layered media.

OCIS codes: 190.0190, 190.6135

    Stable optical solitons have been observed in different nonlocal media, including liquid crystals with reorientational nonlinearity [1,2], and thermal media, where the nonlinear response depends on the boundary conditions [3,4]. The interactions between solitons in such media depend not only on the soliton separation but also on the particular profile of the nonlocal response, thus affording a variety of phenomena that do not occur in materials with local response. Among such phenomena is the formation of multipole soliton states that have been studied in both one- [5-8] and two-dimensional [9-13] geometries.

    Usually, nonlinearity in nonlocal media may be homogeneous across the transverse plane. However, recently structures with spatially inhomogeneous (e.g. periodic) nonlinearities became available too [14-18], thus opening up new possibilities for soliton control. The difference between light propagation in systems with periodic nonlinearity and conventional systems with periodic linear refractive index is crucial, since in the former case light beams self-induce spatially modulated lattices. The effect of a periodic nonlinearity modulation can be more pronounced in nonlocal media than in local materials [14-18], because the entire beam shape and nonlinear response function profile affect the local

nonlinear index contribution. In this Letter we address the formation and propagation of solitons in layered thermal media made of alternating focusing and defocusing layers. We find that in such media bright solitons form in any of the focusing domains, even if they are located near the sample edge and even if the averaged nonlinearity is defocusing.

We consider a laser beam propagating along the $\xi$ axis of a layered thermal medium, where light propagation is described by the system of equations for the dimensionless field amplitude $q$ and normalized temperature variation $T$ [3,11]:

$$
\begin{aligned}
i\frac{\partial q}{\partial \xi} &= -\frac{1}{2}\frac{\partial^2 q}{\partial \eta^2} - \sigma(\eta)qT, \\
\frac{\partial^2 T}{\partial \eta^2} &= -|q|^2.
\end{aligned}
\tag{1}
$$

Here the transverse $\eta$ and longitudinal $\xi$ coordinates are scaled to the beam width and the diffraction length, respectively; the function $\sigma(\eta) = \sigma_{\mathrm{a}} \operatorname{sgn}[\cos(\pi\eta/d)] + \sigma_{\mathrm{b}}$ describes the periodic profile of nonlinearity; $\sigma_{\mathrm{b}}$ is the constant part of nonlinearity; $d$ is the layer width; and the parameter $\sigma_{\mathrm{a}} = \pm 1$ when $\sigma_{\mathrm{b}} = 0$ determines the type of nonlinearity in the central domain (focusing for $\sigma_{\mathrm{a}} = +1$ and defocusing for $\sigma_{\mathrm{a}} = -1$). The nonlinear refractive index change is given by $\delta n = \beta \delta T$, where $\delta T$ is the temperature variation that obeys the steady-state thermo-conductivity equation $\kappa \Delta_\perp \delta T = -\alpha I$, with $\alpha, \beta, \kappa$ being the optical absorption, thermo-optic and thermal conductivity coefficients, respectively, and $I \sim |q|^2$ being the light intensity. We assume that the thermo-optic coefficient $\beta$ changes its sign and/or magnitude in the adjacent layers and that there is no linear refractive index modulation in Eqs. (1). We also assume that the opposite boundaries of the thermal medium $(-L/2, +L/2)$ are thermo-stabilized, and we solved Eqs. (1) with the boundary conditions $q, T|_{\eta \to \pm L/2} = 0$.

We search for solutions of Eqs. (1) in the form $q(\eta, \xi) = w(\eta)\exp(ib\xi)$. To analyze their stability we solved the eigenvalue problem obtained upon substitution of perturbed solutions in the form $q = [w + u\exp(\delta\xi) + iv\exp(\delta\xi)]\exp(ib\xi)$ into Eqs. (1) and subsequent linearization. Namely,

$$\delta u = -\frac{1}{2}\frac{d^2 v}{d\eta^2} - \sigma T v + bv,$$
$$\delta v = \frac{1}{2}\frac{d^2 u}{d\eta^2} + \sigma T u + \sigma \Delta T w - bu,$$
(2)

where $\Delta T = -2\int_{-L/2}^{L/2} G(\eta,\lambda) w(\lambda) u(\lambda) d\lambda$ is the temperature perturbation, and the response function of the thermal medium is given by the expressions $G(\eta,\lambda) = L^{-1}(\eta + L/2)(\lambda - L/2)$ for $\eta \leq \lambda$ and $G(\eta,\lambda) = L^{-1}(\eta - L/2)(\lambda + L/2)$ for $\eta \geq \lambda$. We set a layer width $d = 1$ and a sample width $L = 51.2$. Solitons are characterized by their energy flow $U = \int_{-\infty}^{\infty} w^2 d\eta$.

First we address the case $\sigma_{\rm b} = 0$. Despite the fact that on average the medium exhibits a linear response, one finds a variety of nonlinear solutions including fundamental, even, dipole, and tripole solitons [Figs. 1(a)-1(d)] residing in the center of the sample and supported by self-induced spatially modulated lattice. Importantly, even solitons composed of in-phase spots do not exist in uniform thermal medium and can be found only in layered thermal structures. In all these solutions the growth of the peak intensity is accompanied by light localization in domains with focusing nonlinearity, where $\sigma > 0$, while in the low-power limit solitons expand across several layers and exhibit pronounced amplitude modulations.

Beams propagating in nonuniform thermal media induce spatially modulated lattices that depend on the beam width and peak intensity. Such nonlinear lattices cause the appearance of refractive-index attractors that immobilize soliton and suppress transverse drifts [Fig. 1(f)]. Such attractors allow existence of solitons (including multipoles) not only in the central domain, but also in any focusing domain, even if the domain is located close to the boundary of the sample. While the refractive index profile $\sigma T$ is oscillatory [Fig. 1(f)], its envelope in the narrow region occupied by the laser beam can be approximated by a parabola, a property that explains the net attractive forces acting between out-of-phase spots of multipoles. Far from the region occupied by the laser beam the refractive index envelope is almost linear. Figure 1(e) shows the profile of a shifted fundamental soliton. This is in clear contrast to uniform thermal media where for equal boundary temperatures the only equilibrium point is $\eta = 0$ and all beams launched off-center oscillate along the $\eta$ axis during propagation [4].

The power of the central fundamental soliton vanishes (i.e., $U \to 0$) as $b \to 0$ [Fig. 2(a)], but for shifted solitons centered on focusing domains with positions $|\eta_c| > 0$ there exist a minimal power and a cutoff value of $b$. We set $\eta_c$ as the coordinate of the center of the domain where the soliton peak is located. Thus, for solitons with odd number of spots $\eta_c$ can take values $2nd$ for $\sigma_a > 0$ and $(2n+1)d$ for $\sigma_a < 0$, where $n = 0,1,2,\ldots$. Solitons with $\eta_c = 0$ strongly expand in the low-power limit, but their counterparts with $\eta_c > 0$ remain localized near the cutoff and acquire asymmetric shapes [Fig. 1(e)]. The cutoff $b_{co}$ for odd soliton located at $\eta_c + d$ coincides with the cutoff for even soliton with center at $\eta_c$ and shapes of such solitons become close as $b \to b_{co}$. The $U(b)$ dependencies for higher-order solitons are similar to those for fundamental solitons [Fig. 2(b)]. The cutoffs for existence of fundamental and dipole solitons increase with growth of shift [Fig. 2(c)] and diverge as $|\eta_c| \to L/2$.

We found that the stability of solitons usually depends on the position of their centers. Fundamental solitons are stable for any $b$ and $\eta_c$. Interestingly, even solitons that are unstable in linear lattices can be stable for $0 < b < b_{upp}$, but only for $\eta_c = 0$ [see Fig. 2(d) for the dependence of the real part of the growth rate $\delta_r$ on $b$]. All shifted even solitons are unstable. Dipole solitons feature oscillatory instabilities near cutoff [Fig. 2(e)] and become completely stable above a critical value of the propagation constant. We found that the restriction on the maximal number of poles in stable solitons that holds in uniform thermal medium as well as in some other materials [7] does not hold in thermal media with periodic nonlinearity: Multipole solutions can be stable in such material for any number of spots. All of them become stable above a critical $b$ value. The complexity of the instability domains near cutoff increases with the number of poles in the multipole, but usually for shifted multipoles there remain only a few instability regions [Fig. 2(f)].

Next we address the case $\sigma_b \neq 0$. Stable fundamental, even, and multipole solitons are found even for mean defocusing nonlinearity as soon as $|\sigma_b| < |\sigma_a|$. As $\sigma_b$ decreases the refractive index shape $\sigma T$ becomes asymmetric with respect to $\eta$ axis. At fixed $b$, the nonlinear lattice becomes deeper for smaller $\sigma_b$ values [Fig. 3(b)] a property that causes an increasing soliton localization [Fig. 3(a)]. Even if $\sigma_b > |\sigma_a|$ (in this case the total nonlinearity $\sigma$ is focusing everywhere inside the sample), one still can find robust shifted fundamental and higher-order solitons. Adding nonzero background nonlinearity does not change stability properties qualitatively. Higher-order multipoles remain stable even if total nonlinearity is focusing. Interestingly, for multipoles, increasing $\sigma_\mathrm{b}$ causes an expansion of the instability domains located near the cutoffs [see Fig. 3(c) showing this domain for dipoles], while decreasing $\sigma_\mathrm{b}$ may result in the stabilization of multipoles in their entire existence domain.

Summarizing, we addressed the existence and properties of solitons propagating in thermal nonlinear media, with transversally varying nonlinearity. We found that the modulation of the nonlinearity affords new phenomena, including the possibility of rectilinear propagation of asymmetrically-located solitons.

# References with titles

1. C. Conti, M. Peccianti, and G. Assanto, "Observation of optical spatial solitons in highly nonlocal medium," Phys. Rev. Lett. **92**, 113902 (2004).

2. M. Peccianti, K. A. Brzdakiewicz, and G. Assanto, "Nonlocal spatial soliton interactions in nematic liquid crystals," Opt. Lett. **27**, 1460 (2002).

3. C. Rotschild, O. Cohen, O. Manela, M. Segev, and T. Carmon, "Solitons in nonlinear media with an infinite range of nonlocality: first observation of coherent elliptic solitons and of vortex-ring solitons," Phys. Rev. Lett. **95**, 213904 (2005).

4. B. Alfassi, C. Rotschild, O. Manela, M. Segev, and D. Christodoulides, "Boundary force effects exerted on solitons in highly nonlocal nonlinear media," Opt. Lett. **32**, 154 (2007).

5. I. A. Kolchugina, V. A. Mironov, and A. M. Sergeev, "Structure of steady-state solitons in systems with a nonlocal nonlinearity," JETP Lett. **31**, 304 (1980).

6. X. Hutsebaut, C. Cambournac, M. Haelterman, A. Adamski, and K. Neyts, "Single-component higher-order mode solitons in liquid crystals," Opt. Commun. **233**, 211 (2004).

7. Z. Xu, Y. V. Kartashov, and L. Torner, "Upper threshold for stability of multipole-mode solitons in nonlocal nonlinear media," Opt. Lett. **30**, 3171 (2005).

8. D. Mihalache, D. Mazilu, F. Lederer, L.-C. Crasovan, Y. V. Kartashov, L. Torner, and B. A. Malomed, "Stable solitons of even and odd parities supported by competing nonlocal nonlinearities," Phys. Rev. E **74**, 066614 (2006).

9. S. Lopez-Aguayo, A. S. Desyatnikov, Y. S. Kivshar, S. Skupin, W. Krolikowski, and O. Bang, "Stable rotating dipole solitons in nonlocal optical media," Opt. Lett. **31**, 1100 (2006).

10. Y. V. Kartashov, L. Torner, V. A. Vysloukh, and D. Mihalache, "Multipole vector solitons in nonlocal nonlinear media," Opt. Lett. **31**, 1483 (2006).

11. C. Rotschild, M. Segev, Z. Xu, Y. V. Kartashov, L. Torner, and O. Cohen, "Two-dimensional multipole solitons in nonlocal nonlinear media," Opt. Lett. **31**, 3312 (2006).

12. S. Skupin, O. Bang, D. Edmundson, and W. Krolikowski, "Stability of two-dimensional spatial solitons in nonlocal nonlinear media," Phys. Rev. E **73**, 066603 (2006).
13. A. I. Yakimenko, V. M. Lashkin, O. O. Prikhodko, "Dynamics of two-dimensional coherent structures in nonlocal nonlinear media," Phys. Rev. E **73**, 066605 (2006).
14. H. Sakaguchi and B. A. Malomed, "Matter-wave solitons in nonlinear optical lattices," Phys. Rev. E **72**, 046610 (2005).
15. F. K. Abdullaev and J. Garnier, "Propagation of matter-wave solitons in periodic and random nonlinear potentials," Phys. Rev. A **72**, 061605(R) (2005).
16. G. Fibich, Y. Sivan, and M. I. Weinstein, "Bound states of nonlinear Schrödinger equations with a periodic nonlinear microstructure," Physica D **217**, 31 (2006).
17. Y. V. Bludov and V. V. Konotop, "Localized modes in arrays of boson-fermion mixtures," Phys. Rev. A **74**, 043616 (2006).
18. J. Belmonte-Beitia, V. M. Perez-Garcia, V. Vekslerchik, and P. J. Torres, "Lie symmetries and solitons in nonlinear systems with spatially inhomogeneous nonlinearities," Phys. Rev. Lett. **98**, 064102 (2007).

# References without titles

1. C. Conti, M. Peccianti, and G. Assanto, Phys. Rev. Lett. **92**, 113902 (2004).
2. M. Peccianti, K. A. Brzdakiewicz, and G. Assanto, Opt. Lett. **27**, 1460 (2002).
3. C. Rotschild, O. Cohen, O. Manela, M. Segev, and T. Carmon, Phys. Rev. Lett. **95**, 213904 (2005).
4. B. Alfassi, C. Rotschild, O. Manela, M. Segev, and D. Christodoulides, Opt. Lett. **32**, 154 (2007).
5. I. A. Kolchugina, V. A. Mironov, and A. M. Sergeev, JETP Lett. **31**, 304 (1980).
6. X. Hutsebaut, C. Cambournac, M. Haelterman, A. Adamski, and K. Neyts, Opt. Commun. **233**, 211 (2004).
7. Z. Xu, Y. V. Kartashov, and L. Torner, Opt. Lett. **30**, 3171 (2005).
8. D. Mihalache, D. Mazilu, F. Lederer, L.-C. Crasovan, Y. V. Kartashov, L. Torner, and B. A. Malomed, Phys. Rev. E **74**, 066614 (2006).
9. S. Lopez-Aguayo, A. S. Desyatnikov, Y. S. Kivshar, S. Skupin, W. Krolikowski, and O. Bang, Opt. Lett. **31**, 1100 (2006).
10. Y. V. Kartashov, L. Torner, V. A. Vysloukh, and D. Mihalache, Opt. Lett. **31**, 1483 (2006).
11. C. Rotschild, M. Segev, Z. Xu, Y. V. Kartashov, L. Torner, and O. Cohen, Opt. Lett. **31**, 3312 (2006).
12. S. Skupin, O. Bang, D. Edmundson, and W. Krolikowski, Phys. Rev. E **73**, 066603 (2006).
13. A. I. Yakimenko, V. M. Lashkin, O. O. Prikhodko, Phys. Rev. E **73**, 066605 (2006).
14. H. Sakaguchi and B. A. Malomed, Phys. Rev. E **72**, 046610 (2005).
15. F. K. Abdullaev and J. Garnier, Phys. Rev. A **72**, 061605(R) (2005).
16. G. Fibich, Y. Sivan, and M. I. Weinstein, Physica D **217**, 31 (2006).
17. Y. V. Bludov and V. V. Konotop, Phys. Rev. A **74**, 043616 (2006).
18. J. Belmonte-Beitia, V. M. Perez-Garcia, V. Vekslerchik, and P. J. Torres, Phys. Rev. Lett. **98**, 064102 (2007).

# Figure captions

Figure 1. Profiles of (a) fundamental, (b) even, (c) dipole, and (d) triple-mode solitons for $\eta_c = 0$. (e) Profiles of fundamental solitons for $\eta_c = 10$. (f) Profile of refractive index $\sigma T$ for fundamental solitons with $b = 0.9$, $\eta_c = 0$ (red curve) and $b = 6.5$, $\eta_c = 10$ (black curve). In gray regions $\sigma > 0$, while in white regions $\sigma < 0$. In all cases $\sigma_b = 0$.

Figure 2. $U$ versus $b$ for (a) fundamental and (b) dipole solitons. Circles in (a) and (b) correspond to solitons shown in Figs. 1(a) and 1(c), respectively. (c) Cutoffs for fundamental and dipole solitons versus $\eta_c$. $\delta_r$ versus $b$ for (d) even solitons at $\eta_c = 0$, (e) dipole soliton at $\eta_c = 1$, (f) triple-mode soliton at $\eta_c = 18$. In all cases $\sigma_b = 0$.

Figure 3. (a) Profiles and (b) refractive index distributions for fundamental solitons with $b = 0.5$, $\eta_c = 0$ for $\sigma_b = -0.2$ (red curves) and $\sigma_b = -0.6$ (black curves). (c) Instability domain (shaded) for dipole solitons with $\eta_c = 1$ on the $(\sigma_b, b)$ plane.

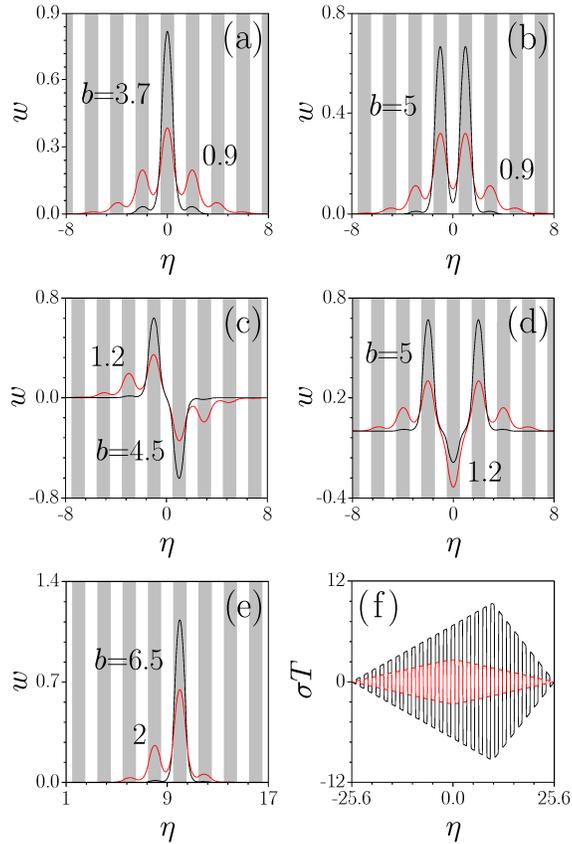

Figure 1. Profiles of (a) fundamental, (b) even, (c) dipole, and (d) triple-mode solitons for $\eta_\mathrm{c} = 0$. (e) Profiles of fundamental solitons for $\eta_\mathrm{c} = 10$. (f) Profile of refractive index $\sigma T$ for fundamental solitons with $b = 0.9$, $\eta_\mathrm{c} = 0$ (red curve) and $b = 6.5$, $\eta_\mathrm{c} = 10$ (black curve). In gray regions $\sigma > 0$, while in white regions $\sigma < 0$. In all cases $\sigma_\mathrm{b} = 0$.

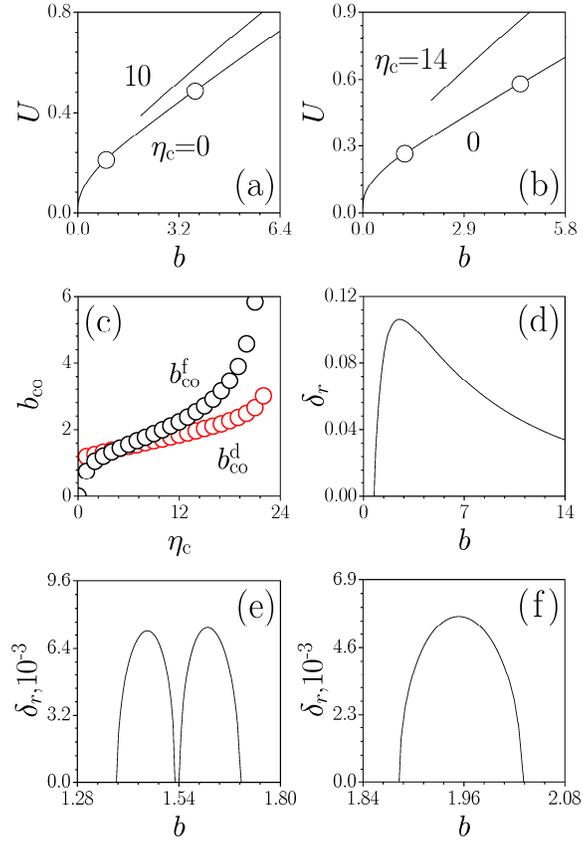

Figure 2. $U$ versus $b$ for (a) fundamental and (b) dipole solitons. Circles in (a) and (b) correspond to solitons shown in Figs. 1(a) and 1(c), respectively. (c) Cutoffs for fundamental and dipole solitons versus $\eta_c$. $\delta_r$ versus $b$ for (d) even solitons at $\eta_c = 0$, (e) dipole soliton at $\eta_c = 1$, (f) triple-mode soliton at $\eta_c = 18$. In all cases $\sigma_b = 0$.

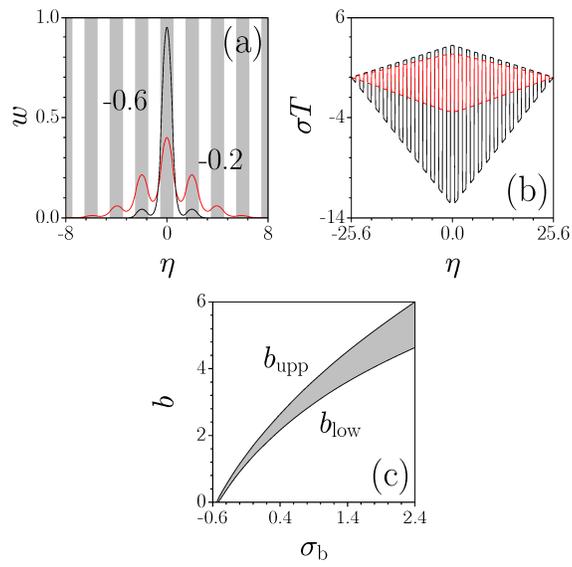

Figure 3. (a) Profiles and (b) refractive index distributions for fundamental solitons with $b = 0.5$, $\eta_c = 0$ for $\sigma_b = -0.2$ (red curves) and $\sigma_b = -0.6$ (black curves). (c) Instability domain (shaded) for dipole solitons with $\eta_c = 1$ on the $(\sigma_b, b)$ plane.

12